\documentclass[%
reprint,
superscriptaddress,
 amsmath,amssymb,
 aps,
pra,
]{revtex4-2}

\usepackage{graphicx}
\usepackage{dcolumn}
\usepackage{bm}
\usepackage{qcircuit}
\usepackage{xcolor}
\usepackage{braket}
\usepackage[counterclockwise, figuresleft]{rotating}

\usepackage{slashbox}
\usepackage{lineno}

\newcommand{\transpose}{\top}
\newcommand{\complexi}{\mathrm{i}}

\begin{document}

\preprint{APS/123-QED}


\title{Quantum estimation bound of Gaussian matrix permanent}

\author{Joonsuk Huh}
 \email{joonsukhuh@gmail.com}
\affiliation{Department of Chemistry, Sungkyunkwan University, Suwon 16419, Korea}
\affiliation{SKKU Advanced Institute of Nanotechnology (SAINT), Sungkyunkwan University, Suwon 16419, Korea}
\affiliation{Institute of Quantum Biophysics, Sungkyunkwan University, Suwon 16419, Korea}

\date{\today}

\begin{abstract}
Exact calculation and even multiplicative error estimation of matrix permanent are challenging for both classical and quantum computers. Regarding the permanents of random Gaussian matrices, the additive error estimation is closely linked to boson sampling, and achieving multiplicative error estimation requires exponentially many samplings. Our newly developed formula for matrix permanents and its corresponding quantum expression have enabled better estimation of the average additive error for random Gaussian matrices compared to Gurvits' classical sampling algorithm. The well-known Ryser formula has been converted into a quantum permanent estimator. When dealing with real random Gaussian square matrices of size $N$, the quantum estimator can approximate the matrix permanent with an additive error smaller than $\epsilon(\sqrt{\mathrm{e}N})^{N}$, where $\epsilon$ is the estimation precision. In contrast, Gurvits' classical sampling algorithm has an estimation error of $\epsilon(2\sqrt{N})^{N}$, which is exponentially larger ($1.2^{N}$) than the quantum method. As expected, the quantum additive error bound fails to reach the multiplicative error bound of $(2\pi N)^{1/4}\epsilon(\sqrt{N/\mathrm{e}})^{N}$. Additionally, the quantum permanent estimator can be up to quadratically faster than the classical estimator when using quantum phase estimation-based amplitude estimation.           
\end{abstract}

\maketitle


\section{Introduction}

Almost three decades ago, Peter Shor~\cite{Shor1999} discovered a celebrated polynomial-time quantum algorithm for integer factorization. Although the best-known classical algorithm for integer factorization scales quasi-exponentially, the problem's computational complexity still needs to be settled. Thus, claiming quantum supremacy with Shor's quantum algorithm is unclear~\cite{Harrow2017}. Moreover, implementing Shor's quantum algorithm for factoring large integer numbers takes a lot of work.

In this context, Aaronson and Arkhipov~\cite{Aaronson2011} proposed a quantum sampling problem known as boson sampling, which has the potential to challenge the extended Church-Turing thesis and demonstrate quantum supremacy. The difficulty of boson sampling is based on the computational complexity of matrix permanents. Approximating permanents of arbitrary complex or real square matrices within a constant factor is a \#P-hard problem. And boson sampling is a scattering process of non-interacting single photons in a linear optical network~\cite{Troyansky1996,Scheel2004}. The output photon distribution of the boson sampling device is expressed as a function of permanents of random Gaussian matrices (RGMs).
Aaronson and Arkhipov demonstrated that if an efficient classical simulation of boson sampling, specifically an additive error matrix permanent estimator, were possible, it would lead to the collapse of the polynomial hierarchy (PH) to the third level, which is widely believed not to be the case. Therefore, the classical hardness of boson sampling relies on the complexity of calculating the permanents of complex RGMs. Still, the \#P-hardness of permanents of Gaussian matrices has yet to be proven that the permanent of Gaussian conjecture (PGC) asserts that the multiplicative error estimation of permanents of RGMs is \#P-hard. Additionally, assuming a non-uniform distribution of permanents of RGMs by the permanent anti-concentration conjecture (PAC), the multiplicative error estimation is essentially equivalent to the additive error estimation within an error proportional to their average magnitude.

Troyansky and Tishby~\cite{Troyansky1996} demonstrated a connection between a Hermitian matrix permanent and the expectation value of a linear optical quantum observable. They found that computing the permanent requires exponentially many samples due to significant observable variance. 
Therefore, the boson sampling device is impractical in estimating permanents, moreover, the magnitudes are generally exponentially small. Although there have been limited attempts to link quantum circuit models to matrix permanents, most studies have focused on proving the computational hardness of computing permanents using quantum circuit models~\cite{Rudolph2009,Aaronson2011hard,Fefferman2017,Park2023}, rather than developing quantum algorithms for estimating matrix permanents.

Even though there is no proof, it is believed that quantum computers cannot solve the \#P-hard problem in polynomial time. It is important to note that estimating the matrix permanent, in this paper, involves additive error approximation, and not exact calculation or relative error estimation, which are widely believed to be impossible even for quantum computers. This paper aims to explore whether quantum computers can estimate matrix permanents and how their performance compares to classical sampling algorithms.

We have developed a new formula to computing an $N$-dimensional matrix permanent based on Ryser's formula~\cite{Ryser1963}. Our formula  expresses the permanent as a sum of $N$-th powers of a binary function, which allows us to reformulate the power expansion as an overlap integral using the quantum Ising Hamiltonian propagator. We have proposed additive error protocols using the Hadamard test~\cite{Aharonov2007,Aharonov2009} and quantum phase estimation (QPE)~\cite{Knill2007}. We found a recently published paper~\cite{Alase2023}, which connects matrix permanents to spin models as Fefferman et al.~\cite{Fefferman2017} and Peropadre et al.~\cite{Peropadre2017}. We compared our results of real Gaussian matrices using the method of Fefferman et al.~\cite{Fefferman2017} to the result of Gurvits' classical sampling algorithm~\cite{gurvits2005,Aaronson2011}, and found an improved upper bound of additive error estimation of Gaussian matrix permanents compared to the existing classical sampling algorithm. However, it cannot achieve the expected relative error limit, as this would lead to the collapse of PH to BQP (bounded-error quantum polynomial time). Additionally, we note that the quantum estimator can achieve a quadratic speedup when using QPE-based amplitude estimation. 

 The paper's structure is as follows: First, we transform the Ryser formula into a sum of $N$-th power of binary functions. Then, we demonstrate that a matrix permanent can be approximately represented as a quantum overlap integral, followed by the additive error analysis with real RGMs. Finally, we discuss the implications of our findings and conclude the paper.
  
\section{Results}
\subsection{Classical formulas}
Recall the definition of permanent of a square matrix $A\in \mathbb{C}^{N\times N}$:
\begin{align}
\mathrm{Per}(A)=\sum_{\vec{{\sigma} }\in S_{\mathrm{p}}}\prod_{j=1}^{N}A_{j,\sigma_{j}} ,
\label{eq:PERA}
\end{align}
where $A_{jk}$ are the entries of $A$ and the set $S_{\mathrm{p}}$ includes all the  permutations of $(1,2,\ldots,N)$, $\{\vec{\sigma} \}$ ($\vec{\sigma} =(\sigma_1,\ldots \sigma_N)$ is an $N$-dimensional vector).  
Brute force computation of a matrix permanent in Eq. \eqref{eq:PERA} requires $N!$ terms in the summation, and each term is composed of products of  $N$ elements of the matrix.
In 1963, Ryser~\cite{Ryser1963} discovered a formula for matrix permanent, which involves only $2^{N}$ terms in the summation,  
\begin{align}
\textrm{Per}(A) = (-1)^{N}\sum_{S\subseteq \{1,\ldots,N\}}(-1)^{\vert S \vert}
\prod_{j=1}^N \sum_{k\in S}A_{jk} ,
\label{eq:Ryser}
\end{align}
where $|S|$ is the cardinality of subset $S$. Although this is a dramatic improvement, the algorithm is still not classically efficient as it scales exponentially. 
We can rewrite the Ryser formula~\eqref{eq:Ryser} by introducing an $N$-dimensional binary vector $\vec{y}$ as follows:  
\begin{align}
\textrm{Per}(A) = \sum_{\vec{y}\in \{0,1\}^{N}}(-1)^{\sum_{k=1}^{N}(y_{k}+1)}
\prod_{j=1}^N (\vec{A}_{j}\cdot\vec{y}) ,
\label{eq:Ryser2}
\end{align}
where $\vec{A}_{j}$ is the $j$-th row of $A$. 
By introducing a variable $x_{k}=2y_{k}-1$, which maps from $\{0,1\}^{N}$ to $\{-1,1\}^{N}$, we can find the Glynn  formula~\cite{Glynn2010,aaronson2012,Chin2018,Yung2019}: 
\begin{align}
\textrm{Per}(A) = \frac{1}{2^{N}}\sum_{\vec{x}\in \{-1,1\}^N}\left(\prod_{k=1}^{N}x_{k}\right)\left(\prod_{j=1}^N\vec{A}_{j}\cdot\vec{x}\right) ,
\label{eq:Glynn}
\end{align}
where we used the identity $\sum_{\vec{x}\in\{-1,1\}^{N}}\prod_{k}^{N}x_{k}=0$. 
Eq.~\ref{eq:Glynn} shows the magnitude is bounded by $\Vert A\Vert^{N}$.  
Gurvits~\cite{gurvits2005,Aaronson2011} suggested an additive error algorithm $\pm\epsilon\Vert A \Vert^{N}$ based on Glynn's formula by interpreting it as an expectation value $E(\prod_{k=1}^{N}x_{k}\prod_{j=1}^N\vec{A}_{j}\cdot\vec{x})$ over the random binary vector $\vec{x}$. $\Vert \cdot \Vert$ is the matrix 2-norm; otherwise, the order is indicated.   

Using the identity of a monomial by Lemma 1 of Kan~\cite{kan:2007}, we have     
\begin{align}
    \prod_{j=1}^{N}\vec{A}_{j}\cdot\vec{x}=\frac{1}{N!2^{N}}\sum_{\vec{x}'\in\{-1,1\}^{N}}\left(\prod_{k=1}^{N}x'_{k}\right)\left(\vec{x'}^{\transpose}A\vec{x}\right)^{N} .
    \label{eq:monomial}
\end{align}
Now, we have a new expression for the matrix permanent by inserting Eq.~\eqref{eq:monomial} into Eq.~\eqref{eq:Glynn} as follows: 
\begin{align}
    \mathrm{Per}(A)=\frac{1}{N!2^{2N}}\sum_{\vec{x},\vec{x}'\in\{-1,1\}^{N}}\left(\prod_{k=1}^{N}x_{k}x_{k}'\right)\left(\vec{x'}^{\transpose}A
    \vec{x}\right)^{N} ,
    \label{eq:Kanexpression}
\end{align} 
whose magnitude is bounded by $\Vert A \Vert_{1,1}^{N}/N!$ and $\Vert A \Vert_{1,1} =\sum_{j,k}\vert A_{jk}\vert$. 

We name our new finding in  Eq.~\eqref{eq:Kanexpression} the Glynn-Kan-Huh (GKH) formula in this paper. 
Thus, Per($A$) can be interpreted as an average of $2^{2N}$ $N$-th power of binary functions by  Eq.~\eqref{eq:Kanexpression}. The implication of the GKH formula is discussed in Appendix A. 

At a glance, our new formula for permanents in   Eq.~\eqref{eq:Kanexpression} is  exponentially more inefficient compared with the original Ryser formula, Eq.~\eqref{eq:Ryser}. However, our new formula is suitable for developing quantum algorithms for permanents because they are composed of an expectation value of $N$-th powers of the binary function, i.e. $\vec{x'}^{\transpose}A
    \vec{x}$. 
In the following section, we develop a quantum algorithm for additive error approximation of matrix permanents. 

\subsection{Quantum estimator}
From the GKH formula in Eq.~\eqref{eq:Kanexpression}, we  immediately identify the following quantum expectation value as a matrix permanent: 
\begin{align}
&\textrm{Per}(A) =\nonumber \\ 
&\tfrac{1}{N!2^{2N}}\sum_{y=0}^{2^{2N}-1}
\langle y\vert\left(\bigotimes_{k=1}^{N}Z_{k}Z_{N+k}\right)
\left(\sum_{j,k=1}^{N}A_{jk}Z_{N+j}Z_{k}\right)^{N}\vert y \rangle ,
\label{eq:GlynnexpX}
\end{align}
where $\vert y\rangle=\vert y_{1}\rangle\cdots\vert y_{2N}\rangle$, $y=y_{1}\cdots y_{2N}$ is given as a $2N$-bit string and $Z_{k}$ is a Pauli-Z operator of the $k$-th qubit. Using the identity, $\frac{1}{2^{N}}\sum_{y=0}^{2^{2N}-1}
\vert y\rangle=\bigotimes_{k=1}^{2N}H_{k}\vert \mathbf{0}\rangle=\vert\phi\rangle$, where $H_{k}$ is a Hadamard operator of the $k$-th qubit and $\vert \mathbf{0}\rangle=\vert 0\rangle^{\otimes 2N}$, we rewrite the matrix permanent as an expectation value of a quantum operator, i.e., $\mathrm{Per}(A)=\frac{1}{N!2^{2N}}\mathrm{Tr}[G_{Z}(A)]=\frac{1}{N!}\langle\phi\vert G_{Z}(A)\vert\phi\rangle=\frac{1}{N!}\langle \mathbf{0}\vert G_{X}(A) \vert \mathbf{0}\rangle$, where $G_{X}(A)=(\bigotimes_{k=1}^{2N}H_{k})G_{Z}(A)(\bigotimes_{k=1}^{2N}H_{k})$ and $X_{k}$=$H_{k}Z_{k}H_{k}$. 
Let us call $G_{X}(A)$ the GKH operator, defined as 
\begin{align}
G_{X}(A)=\left(\bigotimes_{k=1}^{N}X_{k}X_{N+k}\right)
\left(\sum_{j,k=1}^{N}A_{jk}X_{N+j}X_{k}\right)^{N},  
\label{eq:Glynn-KanoperatorX}
\end{align}
which is factored with two components--parity operator and Ising Hamiltonian operator, respectively denoted as $P$ and $\mathcal{H}$. That is, 
$G_{X}(A)=P\mathcal{H}^{N}(A)$, where 
$\mathcal{H}(A)=\sum_{j,k=1}^{N}A_{jk}X_{N+j}X_{k}$ and 
$P=\bigotimes_{k=1}^{N}X_{k}X_{N+k}$.
Thus far, we have expressed the matrix permanent as an expectation value of qubit operators and qubit states. The resulting expression involves a power operator of $\mathcal{H}$, which is non-unitary and generally non-Hermitian. The remaining task is to approximately implement the power operator in a quantum circuit. We can perform the approximation through various  approaches~\cite{Aulicino2022}. For example, we can consider quantum signal processing with quantum singular value transformation and the linear combination of unitaries (LCU) technique~\cite{Childs2012,Berry2015,Low2019}. However, the repeated application of LCU for the power operator requires a prefactor for the normalization of the quantum states grows as the power of the Hamiltonian norm,  $\Vert\mathcal{H}(A)\Vert^{N}$ ($\Vert\mathcal{H}(A)\Vert\le\Vert A\Vert_{1,1}$), resulting in large errors in the end. The quantum power method of Seki et al.~\cite{Seki2021} can also be used. The method is based on the finite difference to approximate the Hamiltonian power operator as the high-order time differentiation of the time propagator. However, it also suffers from a large error due to a convergence condition related to the Hamiltonian norm. 
Instead, we adopt the method by Fefferman et al.~\cite{Fefferman2017}, where the hardness of the exact boson sampling was discussed with permanents of integer matrices because it produces less error than the quantum power method. The matrix permanent is approximated as a single overlap integral of the time-propagator of quantum Ising Hamiltonian. The overlap integral is evaluated within an additive error using the Hadamard test or the QPE-based amplitude estimation. 

\subsubsection{Matrix permanent and time-propagator}
Let's rewrite the matrix permanent with the GKH operator and leave only the Hamiltonian power operator as follows, 
\begin{align}
\mathrm{Per}(A)
&=\frac{1}{N!}\langle \mathbf{0}\vert P\mathcal{H}(A)^{N} \vert \mathbf{0}\rangle  \nonumber \\
&=\langle \mathbf{1}\vert \frac{\mathcal{H}(A)^{N}}{N!} \vert \mathbf{0}\rangle , 
\label{eq:permanentquantumpower}
\end{align}
where $P\vert\mathbf{0}\rangle=\vert\mathbf{1}\rangle=\vert 1\cdots 1\rangle$. 
Eq.~\eqref{eq:permanentquantumpower} can be interpreted as a transition amplitude from $\vert\mathbf{0}\rangle$ to $\vert\mathbf{1}\rangle$. A single application of $\mathcal{H}(A)$ results in the pairwise bit flips with amplitudes $A_{jk}$, and thus, the matrix permanent is a counting amplitudes problem.  
On the other hand, we can consider an overlap integral of the time-propagator, i.e.,  
\begin{align}
\langle \mathbf{1}\vert \exp (-\complexi \mathcal{H}(A) t) \vert \mathbf{0}\rangle 
=\langle \mathbf{1}\vert \sum_{l=0}^{\infty}\frac{(-\complexi t\mathcal{H}(A))^{N+2l}}{(N+2l)!} \vert \mathbf{0}\rangle ,
\end{align}
where the terms of Taylor series in a lower order than $N$ vanish because the scattering operations of $\mathcal{H}(A)$ on $\vert\mathbf{0}\rangle$ at least $N$ times to form $\vert\mathbf{1}\rangle$ and the pairwise selection should be made an even number of times for the amplitude to survive that the additional odd order (from the $N$-th order) contributions vanish~\cite{Fefferman2017}. Now, we can relate the overlap integral of the time-propagator with the matrix permanent as follows: 
\begin{align}
\langle \mathbf{1}\vert \exp (-\complexi \mathcal{H}(A) t) \vert \mathbf{0}\rangle 
&=(-\complexi t)^{N}\mathrm{Per}(A)+ \sum_{l=1}^{\infty}(-\complexi t)^{N+2l}R_{l} ,
\label{eq:expamplitude}
\end{align}
where the overlap integral is expanded as a time-series that the leading order's coefficient is the matrix permanent, and the remaining error terms follow: $R_{l}=\langle \mathbf{1}\vert
    \frac{\mathcal{H}(A)^{N+2l}}{(N+2l)!} \vert \mathbf{0}\rangle$. 
    Here, we assume $A$ is a real matrix to have the time-propagator as a unitary operator, and $N$ is an even number for brevity. So only the real part matters. For odd $N$, we need to take care of the imaginary part instead of the real part.     
After rearranging Eq.~\eqref{eq:expamplitude}, we can express the matrix permanent as follows: 
    \begin{align}
\mathrm{Per}(A)=(-t^{2})^{-N/2}\left[\langle \mathbf{1}\vert \exp (-\complexi \mathcal{H}(A) t) \vert \mathbf{0}\rangle 
+R\right] ,
\label{eq:expamplitudeforper}
\end{align}
where $R=-\sum_{l=1}^{\infty}(-t^{2})^{N/2+l}R_{l}$ and the equation implies if the remaining error term is small enough, we can approximate the matrix permanent via estimating the overlap integral of the time-propagator. Note that we only need the real part of the overlap integral because only the real part survives. The quantum estimation of the overlap integral can be made by the Hadamard test~\cite{Aharonov2007,Matsuo2014,Huang2019} or the QPE-based method~\cite{Knill2007}, whose costs scale $\mathcal{O}(1/\epsilon^{'2})$ and $\mathcal{O}(1/\epsilon')$, respectively. 
We note here that the computational cost of the Hadamard test of the Ising Hamiltonian scales identically to the classical samplings because the Hamiltonian is provided in a computational basis. Thus, the computational advantage can be achieved only in the QPE approach, which scales quadratically better.

Now the matrix permanent estimation problem becomes an evaluation problem of the real part of the quantum overlap integral, i.e., $\mathrm{Re}\braket{\phi\vert U \vert \phi}=\mathrm{Re}\braket{\mathbf{0}\vert (\bigotimes_{k=1}^{2N}H_{k})U (\bigotimes_{k=1}^{2N}H_{k})\vert \mathbf{0}}$, for a unitary operator $U$, which includes the parity operator and the time-propagator.  
We present a Hadamard test quantum circuit for the overlap integral evaluation in Appendix B. The Hadamard test circuit requires a polynomial number of quantum gates and depth. We refer readers to Knill et al.~\cite{Knill2007} for the QPE-based algorithm. 

\subsubsection{Average additive error upper bound}

This paper focuses on the real RGMs to discuss the algorithm's remarkable implications. 
The average upper bound of $\vert R_{l}\vert$ in Eq.~\eqref{eq:expamplitudeforper} (i.e., $\mathcal{E}( \vert R_{l}\vert$) with respect to real RGMs $\{W\}$, where $W\sim\mathcal{N}(0,1)^{N\times N}_{\mathbb{R}}$ of \emph{i.i.d.} real Gaussians is 
\begin{align}
     \mathcal{E}( \vert R_{l}\vert)&\le
     \frac{1}{2}\mathcal{E}\left(\langle \mathbf{0}\vert
    \frac{\mathcal{H}(W)^{N+2l}}{(N+2l)!} \vert \mathbf{0}\rangle+
    \langle \mathbf{1}\vert
    \frac{\mathcal{H}(W)^{N+2l}}{(N+2l)!} \vert \mathbf{1}\rangle\right) \nonumber\\
    &=\frac{1}{2(N+2l)!}\frac{\partial^{N+2l}}{\partial\beta^{N+2l}}
    \mathcal{E}(\langle \mathbf{0}\vert
    \exp(\beta\mathcal{H}(W)) \vert \mathbf{0}\rangle\nonumber \\
    &~~~~~~~~~~~~~~~~~~~~~~~~~~~~~+\langle \mathbf{1}\vert
    \exp(\beta\mathcal{H}(W)) \vert \mathbf{1}\rangle)\Big\vert_{\beta=0} ,
\end{align}
where we used the fact that $\mathcal{H}^{N+2l}$ is a positive semidefinite operator (by assuming $N$ as an even number) for the inequality and we introduced a generating function with a parameter $\beta$ to perform the Gaussian integral for averaging $\mathcal{E}$ with respect to the Gaussian distribution. The resulting identity is a simple exponential function of $\beta^{2}$: 
\begin{align}
\mathcal{E}\left(\langle \mathbf{0}\vert
    \exp(\beta\mathcal{H}(W)) \vert \mathbf{0}\rangle\right)&=
    \langle \mathbf{0}\vert
    \mathcal{E}\left(
\mathrm{e}^{\beta\sum_{j,k=1}^{N}W_{jk}X_{N+j}X_{k}} \right)\vert \mathbf{0}\rangle \nonumber\\
&=\exp\left(\frac{\beta^{2}N^{2}}{2}\right)\nonumber \\
&=\mathcal{E}\left(\langle \mathbf{1}\vert
    \exp(\beta\mathcal{H}(W)) \vert \mathbf{1}\rangle\right), 
\label{eq:proofgaussianaverage}
\end{align}
where, in the second line, the one-dimensional Gaussian integrals were performed regarding the corresponding normal distributions of $\{W_{jk}\}$, and we used $X_{N+j}^{2}X_{k}^{2}=1$. An alternative proof of the above relation is given in Appendix C.  
Finally, we have an upper bound of $\mathcal{E}( \vert R_{l}\vert)$: 
\begin{align}
    \mathcal{E}( \vert R_{l}\vert)&\le\frac{1}{(N+2l)!}\frac{\partial^{N+2l}}{\partial\beta^{N+2l}}
    \exp\left(\frac{\beta^{2}N^{2}}{2}\right)\Big\vert_{\beta=0} \nonumber\\
    &=\frac{(N^{2}/2)^{N/2+l}}{(N/2+l)!} .
\end{align}
Consequently, the remaining error term in Eq.~\eqref{eq:expamplitudeforper} has an averaged upper bound, 
\begin{align}
\mathcal{E}\left(\left\vert R\right\vert\right)
&\le(Nt^{2})^{N/2}\sum_{l=1}^{\infty}\frac{(N/2)^{N/2+l}(Nt^{2})^{l}}{(N/2+l)!} \nonumber\\
&\le 
(Nt^{2})^{N/2}\left(\sum_{l=1}^{\infty}\frac{(N/2)^{N/2+l}}{(N/2+l)!}\right)\left(\sum_{l=1}^{\infty}(Nt^{2})^{l}\right) \nonumber\\
&< (Nt^{2})^{N/2}\exp(N/2)\left(\sum_{l=1}^{\infty}(Nt^{2})^{l}\right) \nonumber\\
&=\frac{Nt^{2}}{(1-Nt^{2})}
(\mathrm{e}Nt^{2})^{N/2},
\label{eq:avgbound}
\end{align}
where we assumed $Nt^{2}<1$ for the convergence and used 
\begin{align}
\left(\sum_{l=1}^{\infty}\frac{(N/2)^{N/2+l}}{(N/2+l)!}\right)
= \exp(N/2)-\sum_{l=1}^{N/2}\frac{(N/2)^{l}}{l!} .
\label{eq:seriessummation}
\end{align} 
To make the averaged remaining error bounded by an exponentially small number  ($\epsilon_{\mathrm{R}}=(\mathrm{e}-c)^{-1}c^{N/2+1}$), we need  $c=\mathrm{e}Nt^{2}< 1$, which is tighter than $Nt^{2}<1$. 

The quantum amplitude is estimated using the Hadamard test or the QPE-based amplitude estimation algorithm with the additive error precision $\epsilon'$. 
Therefore, the total error $\epsilon_{\mathrm{T}}$ in Eq.~\eqref{eq:expamplitudeforper} has an average upper bound as follows: 
\begin{align}
    \epsilon_{\mathrm{T}}&\le 
t^{-N}\left(\epsilon'+\epsilon_{\mathrm{R}}\right), 
\label{eq:totalupperbound}
\end{align}
where we set $\epsilon=\epsilon'+\epsilon_{\mathrm{R}}$ and the upper bound becomes $\epsilon (\sqrt{\mathrm{e}N/c})^{N}$ when $t=\sqrt{c/(N\mathrm{e})}$. Note that $\epsilon'$ is polynomially small and $\epsilon_{\mathrm{R}}$ is exponentially small. Here, we also note that the quantum algorithm's sampling cost can scale $\mathcal{O}(1/\epsilon')$, which is quadratically better than the classical sampler if we use the QPE-based amplitude estimation~\cite{Knill2007} instead of the Hadamard test. 
Now, we compare our error bound with the Gurvits classical sampling algorithm~\cite{gurvits2005,Aaronson2011} for real RGMs. 
 
\emph{Gurvits algorithm~\cite{gurvits2005,Aaronson2011}--} The Gurvits classical sampling algorithm generates an additive error $\epsilon\Vert W \Vert^{N}$ with a sampling cost $\mathcal{O}(1/\epsilon^{2})$ and the average error is $\epsilon(2\sqrt{N})^{N}$, where we used $\mathcal{E}(\Vert W \Vert)=2\sqrt{N}$~\cite{Aaronson2011,vershynin_2018,Rebrova2018}. This error is an exponential factor $(2/\sqrt{\mathrm{e}/c})^{N}< 1.2^{N}$ (where $\mathrm{e}/4< c<1$) larger than the quantum version. 

\subsubsection{Tighter average error bound using GKH formula}
The upper bound in Eq.~\eqref{eq:avgbound} can be improved by taking the alternating sign changes in the summation into account as follows: 
\begin{align}
    \vert R\vert
    &=\left\vert -\sum_{k=0}^{\infty}
    \left[(-t^{2})^{N/2+2k+1}R_{2k+1}+(-t^{2})^{N/2+2k+2}R_{2k+2}\right]\right\vert \nonumber \\
    &=\left\vert -(-t^{2})^{N/2}\sum_{k=0}^{\infty}(-t^{2})^{2k+1}(R_{2k+1}-t^{2}R_{2k+2})\right\vert. 
\end{align}
Note that $R_{l}\ge0$ ($R_{l}<0$) for all $l$ when $\mathrm{Per}(A)\ge 0$ ($\mathrm{Per}(A)< 0$); this is clearly seen when the structure of $R_{l}$ in eigenvalues is considered. Without losing the generality, we assume $\mathrm{Per}(A)\ge 0$ and thus $R_{l}\ge0$ for all $l$. Similar to the GKH formula~\eqref{eq:Kanexpression}, we can express $R_{l}$ as follows: 
\begin{align}
   R_{l}=\frac{1}{(N+2l)!2^{2N}}\sum_{\vec{x},\vec{x}'\in\{-1,1\}^{N}}\left(\prod_{k=1}^{N}x_{k}x_{k}'\right)\left(\vec{x'}^{\transpose}A
    \vec{x}\right)^{N+2l} .
\end{align} 
Accordingly, $R_l$ can be rewritten as a difference between two positive semidefinite terms, i.e.,  
\begin{align}
    R_{l}&=\sum_{\{\vec{x},\vec{x'}\}\in \{-1,1\}^{N}}\frac{\mathcal{S}_{+}
    \left(\vec{x'}^{\transpose}A
    \vec{x}\right)^{N+2l}}{(N+2l)!2^{2N}}
    -\frac{\mathcal{S}_{-}
    \left(\vec{x'}^{\transpose}A
    \vec{x}\right)^{N+2l}}{(N+2l)!2^{2N}}
    \nonumber \\
    &=R_{l}^{(+)}-R_{l}^{(-)}, 
\end{align}
where the first and second terms are denoted $R_{l}^{(\pm)}$ and $\mathcal{S}_{\pm}$ refer to  $\left(\prod_{k=1}^{N}x_{k} \pm \prod_{k=1}^{N}x_{k}'\right)^{2}/4$, respectively. 
With the new notation, we can recast $R$ as follows: 
\begin{align}
    \vert R\vert=\Big\vert -(-t^{2})^{N/2}\sum_{k=0}^{\infty}(-t^{2})^{2k+1}&[R_{2k+1}^{(+)}-t^{2}R_{2k+2}^{(+)}\nonumber\\
    -(&R_{2k+1}^{(-)}-t^{2}R_{2k+2}^{(-)})]\Big\vert.
\end{align}
When $t\le N$, $R_{2k+1}^{(\pm)}\ge t^{2}R_{2k+2}^{(\pm)}$ because  $0\le R_{2k+1}^{(\pm)}\le R_{2k+2}^{(\pm)}(N+4k+4)(N+4k+3)$, which can be shown by the power mean inequality:  
\begin{align}
    &R_{l+1}^{\pm}(N+2l+2)!\ge [R_{l}^{\pm}(N+2l)!]^{\frac{N+2l+2}{N+2l}}\ge R_{l}^{\pm}(N+2l)!.
\end{align}
The last inequality is generally true for random Gaussian matrices because $\Vert A\Vert_{1,1}=\mathcal{O}(N^{2})$ that $R_{l}^{\pm}(N+2l)!\ge 1$. 
Then, we conclude as follows:   
\begin{align}
R_{l+1}^{\pm}(N+2l+2)(N+2l+1)\ge R_{l}^{\pm}.
\end{align}
Thus, with a small $t$ ($\le N$),  
$R$ is bounded as follows:  
\begin{align}
    \vert R\vert&\le (t^{2})^{N/2}\sum_{k=0}^{\infty}(t^{2})^{2k+1}[R_{2k+1}^{(+)}-t^{2}R_{2k+2}^{(+)}\nonumber \\
    &~~~~~~~~~~~~~~~~~~~~~~~~~~~~~~+(R_{2k+1}^{(-)}-t^{2}R_{2k+2}^{(-)})]\nonumber \\
    &=-(t^{2})^{N/2}\sum_{l=1}^{\infty}(-t^{2})^{l}(R_{l}^{(+)}+R_{l}^{(-)}).
\end{align}
Now we can set the average upper bound of $R$, i.e., 
\begin{align}
    \mathcal{E}(\vert R\vert)&\le-(t^{2})^{N/2}\sum_{l=1}^{\infty}(-t^{2})^{l}\mathcal{E}(R_{l}^{(+)}+R_{l}^{(-)})\nonumber \\
    &=-(-1)^{N/2}\sum_{l=1}^{\infty}\frac{(-N^{2}t^2/2)^{N/2+l}}{(N/2+l)!}\nonumber \\
    &=\left\vert\exp(-N^{2}t^{2}/2)-\sum_{l=0}^{N/2}\frac{(-N^{2}t^2/2)^{l}}{l!}\right\vert\nonumber\\
    &=\frac{\exp(-\theta N^{2}t^{2}/2)}{(N/2+1)!}\left(\frac{N^{2}t^{2}}{2}\right)^{N/2+1}, 
\end{align}
where $0\le\theta<1$ in the Lagrange form of the remainder in Taylor's theorem for the Gaussian function.  
We used the Stirling approximation, 
\begin{align}
\frac{1}{(N/2+1)!}<\frac{1}{\sqrt{2\pi(N/2+1)}}\left(\frac{\mathrm{e}}{N/2+1}\right)^{N/2+1} ,   
\end{align}
for the following inequality, 
\begin{align}
    \frac{\mathrm{e}^{-\theta N^{2}t^{2}/2}\left(\frac{N^{2}t^{2}}{2}\right)^{N/2+1}}{(N/2+1)!}&<
    \frac{\mathrm{e}^{-\theta N^{2}t^{2}/2}\left(\frac{\mathrm{e}N^{2}t^{2}}{N+2}\right)^{N/2+1}}{\sqrt{2\pi(N/2+1)}}
    \nonumber \\
    &=\frac{\left[\exp\left(1-\theta \tau\right)\tau\right]^{N/2+1}}{\sqrt{2\pi(N/2+1)}},
\end{align}
where $\tau=N^{2}t^{2}/(N+2)$. 
To make the average upper bound of $\vert R\vert$ exponentially small, we set $\exp\left(1-\theta \tau\right)\tau<1$ that $\frac{\mathrm{ln}(\mathrm{e}\tau)}{\tau}<\theta$. Therefore, when $\tau=1/\mathrm{e}$, the constant $\theta$ always exists in $(0,1)$. Finally, we have an exponentially small upper bound, 
\begin{align}
    \mathcal{E}(R)<\frac{\mathrm{e}^{-\tfrac{\theta}{\mathrm{e}}}}{\sqrt{2\pi (N/2+1)}}\exp\left(-\tfrac{\theta}{2\mathrm{e}}N\right)=\epsilon_{\mathrm{R}}^{\mathrm{tight}},
\end{align}
with 
\begin{align}
t=\left(\sqrt{\mathrm{e}\left(\frac{N}{N+2}\right)N}\right)^{-1}, 
    \end{align}
    and 
    \begin{align}
        t^{-N}=\left(\sqrt{\mathrm{e}\left(\frac{N}{N+2}\right)N}\right)^{N} .
    \end{align}
Consequently, we have the total error bound as 
\begin{align}
 \epsilon_{\mathrm{T}}&\le 
t^{-N}\left(\epsilon'+\epsilon_{\mathrm{R}}^{\mathrm{tight}}\right)\nonumber \\
&=\epsilon\left(\sqrt{\mathrm{e}\left(\frac{N}{N+2}\right)N}\right)^{N}\nonumber \\
&<\epsilon\left(\sqrt{\mathrm{e}N}\right)^{N},    
\end{align}
where we set $\epsilon=\epsilon'+\epsilon_{\mathrm{R}}^{\mathrm{tight}}$.

\section{Conclusion and Outlook}
We have introduced a new classical formula for matrix permanent called the GKH (Glynn-Kan-Huh) formula. This formula was derived by combining the monomial expansion of Kan ~\cite{kan:2007} with the Glynn formula~\cite{Glynn2010}. Recently, it was rederived using quantum optical language~\cite{Chabaud2022}. While the GKH formula is more inefficient than Ryser's formula, it has a simple structure of the $N$-th power of a binary function, allowing for the potential use of matrix symmetry and linear algebra techniques, such as singular value decomposition, to reduce computational costs. Unlike the Ryser and Glynn formulas, linear algebra can be introduced to the GKH formula. For instance, it is known that the multiplicative error estimations of a 0-1 entry matrix and a Hermitian positive semidefinite matrix can be obtained in polynomial time. The GKH formula could facilitate the development of the BPP$^\mathrm{NP}$ algorithm for the Hermitian positive semidefinite matrix. It is also noteworthy that the GKH formula can be extended to matrices with repeated rows and columns~\cite{aaronson2012,Chin2018,Yung2019,Chabaud2022} and can be used to develop corresponding quantum formulas.

The classical GKH formula can be converted to the quantum expectation value of the GKH operator, allowing for the estimation of matrix permanents using quantum amplitudes. Quantum algorithms based on the GKH operator can take different forms depending on how the Hamiltonian power operator is approximated. The Hamiltonian power operator can be incorporated into a large unitary operator, which may include ancilla qubits, such as LCU or quantum singular value transformation. In either case, the permanent can be expressed as a single overlap integral (denoted as $a$) of the large unitary operator. The resulting permanent expression contains a significant prefactor, i.e., $\mathrm{Per}(A)=\Vert A\Vert^{N}_{1,1}/N!\times a$, resulting in generally superexponentially large (for $\Vert A\Vert_{1,1}\ge\mathcal{O}(N^{2})$) additive error in the additive error estimation protocol (e.g., Hadamard test) for the overlap integral or a multiplicative error with exponentially expensive cost due to the exponentially small amplitude~\cite{Knill2007}. However, it is unknown whether some matrices give polynomially small amplitudes for efficient multiplicative error quantum amplitude estimation~\cite{Knill2007}. When $\Vert A\Vert_{1,1}=\mathcal{O}(N)$ and the size of permanent is only exponentially large, the amplitude can be in a polynomial size. If such matrices exist, the relative error permanent estimation of such matrices may not be \#P-hard, because we strongly believe BQP cannot access the \#P-hardness~\cite{Aaronson2011}; the possibility is still open, but, so far, it was shown by Lloyd et al. ~\cite{Abrams1998} that it requires nonlinearity for a quantum device to access a \#P-hard problem efficiently.

Instead, in this paper, we approximated the matrix permanent by representing it as an overlap integral of a time-propagator of the quantum Ising Hamiltonian. The first surviving term in the Taylor expansion with respect to time $t$ is proportional to the matrix permanent, while the higher-order terms are considered as an additive error. In this paper, we examine the real RGMs and discuss the implications of our findings. By imposing a condition on the size of the time parameter, $t=\mathcal{O}(1/\sqrt{N})$, we showed that the averaged remaining error can be exponentially small. As a result, we can approximate the permanents of real RGMs with an average additive error smaller than $\epsilon (\sqrt{\mathrm{e}N})^{N}$, which represents an exponential factor ($1.2^{N}$) improvement compared to the Gurvits classical sampler. It is important to note that our upper bound might be more optimal, and there could be potential for improving the quantum estimator's error. However, we caution that achieving the relative error limit ($(2\pi N)^{1/4}\epsilon(\sqrt{N/\mathrm{e}})^{N}$) would potentially lead the Polynomial Hierarchy to collapse to BQP, due to conjectures (PGA and PAC)~\cite{Aaronson2011}.

Interestingly, the complexity of the circuit depends on the matrix elements of the quantum Ising Hamiltonian, as different matrices result in various circuit complexities~\cite{Matsuo2014,Fujii2017,Mann2019}. Thus, the quantum circuit model can be used to analyze the computational complexity of a matrix permanent (\emph{cf.}~\cite{Park2023}).

Although the estimation error of the quantum estimator did not meet the relative error limit of $(2\pi N)^{1/4}\epsilon(\sqrt{N/\mathrm{e}})^{N}$, our quantum analysis represents a significant breakthrough in the quantum estimation of matrix functions such as (loop)-hafnians, Torontonians, determinants, and immanants with minor modifications. More details about this work will be available soon.

\begin{acknowledgments}
The author acknowledges Raymond Kan, Hyukjoon Kwon, Dominik Hangleiter, Gwonhak Lee, Chae-Yeun Park, and Taewan Kim for the fruitful discussions. 
This work was supported by Basic Science Research Program through the National Research Foundation of Korea (NRF), funded by the Ministry of Education, Science and Technology (NRF-2021M3E4A1038308, NRF-2021M3H3A1038085, NRF-2019M3E4A1079666, NRF-2022M3H3A106307411, NRF-2023M3K5A1094805, NRF-2023M3K5A1094813). 
This work was also partly supported by Institute for Information \& communications Technology Promotion (IITP) grant funded by the Korea government(MSIP) (No. 2019-0-00003, Research and Development of Core technologies for Programming, Running, Implementing and Validating of Fault-Tolerant Quantum Computing System). 
\end{acknowledgments}

\section*{Appendix}
\appendix

\section{Classical implication of of the GKH formula (Eq.~\eqref{eq:Kanexpression})}
Interestingly, Eq.~\eqref{eq:Kanexpression} can be rewritten as a difference between two positive semidefinite functions for a real square matrix of $A$ with an even dimension, 
\begin{align}
    \mathrm{Per}(A)=&\frac{1}{N!2^{2N}}\Huge[\sum_{\{\vec{x},\vec{x'}\}\in \{-1,1\}^{N}}
    \mathcal{S}_{+}
    \left(\vec{x'}^{\transpose}A
    \vec{x}\right)^{N}\nonumber\\
    &-\sum_{\{\vec{x},\vec{x'}\}\in \{-1,1\}^{N}}
    \mathcal{S}_{-}
    \left(\vec{x'}^{\transpose}A
    \vec{x}\right)^{N}\Huge], 
    \label{eq:gapp}
\end{align}
where $\mathcal{S}_{\pm}$ refers to  $\left(\prod_{k=1}^{N}x_{k} \pm \prod_{k=1}^{N}x_{k}'\right)^{2}/4$, respectively. For the odd-dimensional case, we can use the following identity to have the even-dimensional matrix, $\mathrm{Per}(A)= \mathrm{Per}\left(A\bigoplus 1\right)$. 
Eq.~\eqref{eq:gapp} directly shows \#P-hard is GapP-hard by the definition of GapP~\cite{Hangleiter}. On the other hand, the multiplicative error estimation of a permanent of real matrix is also \#P-hard. The positive semidefinite functions in Eq.~\eqref{eq:gapp} can be evaluated with multiplicative error estimation using Stockmeyer's BPP$^{\mathrm{NP}}$ counting algorithm~\cite{Aaronson2011,Stockmeyer1985}, where BPP and NP stand for bounded error probabilistic polynomial time and nondeterministic polynomial time, respectively.

Meanwhile, the multiplicative error estimation of the permanent of a Hermitian positive semidefinite matrix is classified as   $\mathrm{BPP^{NP}}$~\cite{Lund2014,Rahimi-Keshari2015,Chakhmakhchyan2017}. Although it is known that permanents of non-negative matrices can be evaluated within a constant factor by Jerrum-Sinclair-Vigoda (JSV) algorithm~\cite{Jerrum2004} in polynomial time, the JSV algorithm has no practical implementations so far~\cite{Bezakova2008,Newman2020}. 
Therefore, further computational complexity analysis of estimating permanents of real matrices using Eq.~\eqref{eq:gapp} is required. We would be able to discover new efficient classical algorithms with the GKH formula.

On the other hand, we have a condition for the superiority of the GKH sampler (Monte Carlo sampling with the GKH formula) with respect to the Gurvits algorithm as  \begin{align}
    \epsilon\frac{\Vert A\Vert_{1,1}^{N}}{N!}\le \epsilon\Vert A \Vert^{N}, 
    \label{eq:betterboundoriginal}
\end{align}
which implies 
\begin{align}
    \Vert A\Vert_{1,1}\le \frac{N}{\mathrm{e}}\Vert A \Vert,
    \label{eq:betterbound}
\end{align}
where we used $N!\simeq (N/\mathrm{e})^{N}$. 

Based on this condition~\eqref{eq:betterbound}, we may consider three cases:  
\begin{enumerate}
    \item $\Vert A\Vert_{1,1}\le\frac{N}{e}\Vert A \Vert\le\frac{N}{e}$, which implies $\Vert A \Vert \le 1$: The GKH sampler produces exponentially small errors, smaller than Gurvits' exponentially small error. 
    \item $\Vert A\Vert_{1,1}\le\frac{N}{e}\le\frac{N}{e}\Vert A \Vert$, which implies $\Vert A \Vert \ge 1$: The GKH sampler still produces exponentially small errors, but the Gurvits algorithm makes an exponentially large error.
     \item $\frac{N}{e}\le\Vert A\Vert_{1,1}\le\frac{N}{e}\Vert A \Vert$, which implies $\Vert A \Vert \ge 1$: Both classical samplers produce exponentially large errors, but the GKH sampler's error is smaller than the Gurvits' one.  
\end{enumerate}

We can set a tighter condition with the 1-norm and 2-norm relation as 
\begin{align}
    \Vert A\Vert_{1,1}\le N\Vert A \Vert_{1} \le\frac{N}{e}\Vert A \Vert,
\end{align}
where we set the upper bound, as in Eq.~\eqref{eq:betterbound}, to find a sufficient condition of matrix norms. 
Recalling the well-known inequality, $1/\sqrt{N}\le\Vert A\Vert/\Vert A\Vert_{1}\le\sqrt{N}$, we have
\begin{align}
    e\le\frac{\Vert A\Vert}{\Vert A\Vert_{1}}\le \sqrt{N}, 
    \label{eq:ratiobound}
\end{align}
which implies $e^{2}\le N$. We observe that the domain in Eq.~\eqref{eq:ratiobound} is non-empty for the sufficiently large $N\ge 8$ and even it is dominant (50 \%) when $N\ge 28$. However, we note, here, that the random Gaussian matrices do not belong to the domain in average because the average 1-norm and 2-norm are $\mathcal{O}(N)$ and $\mathcal{O}(\sqrt{N})$, respectively. Therefore, the Gurvits sampler performs better than the GKH sampler for random Gaussian matrices.    

\section{Circuit synthesis for Hadamard test}
\begin{figure}[htb]
\centering
    \includegraphics[width=0.5\textwidth]{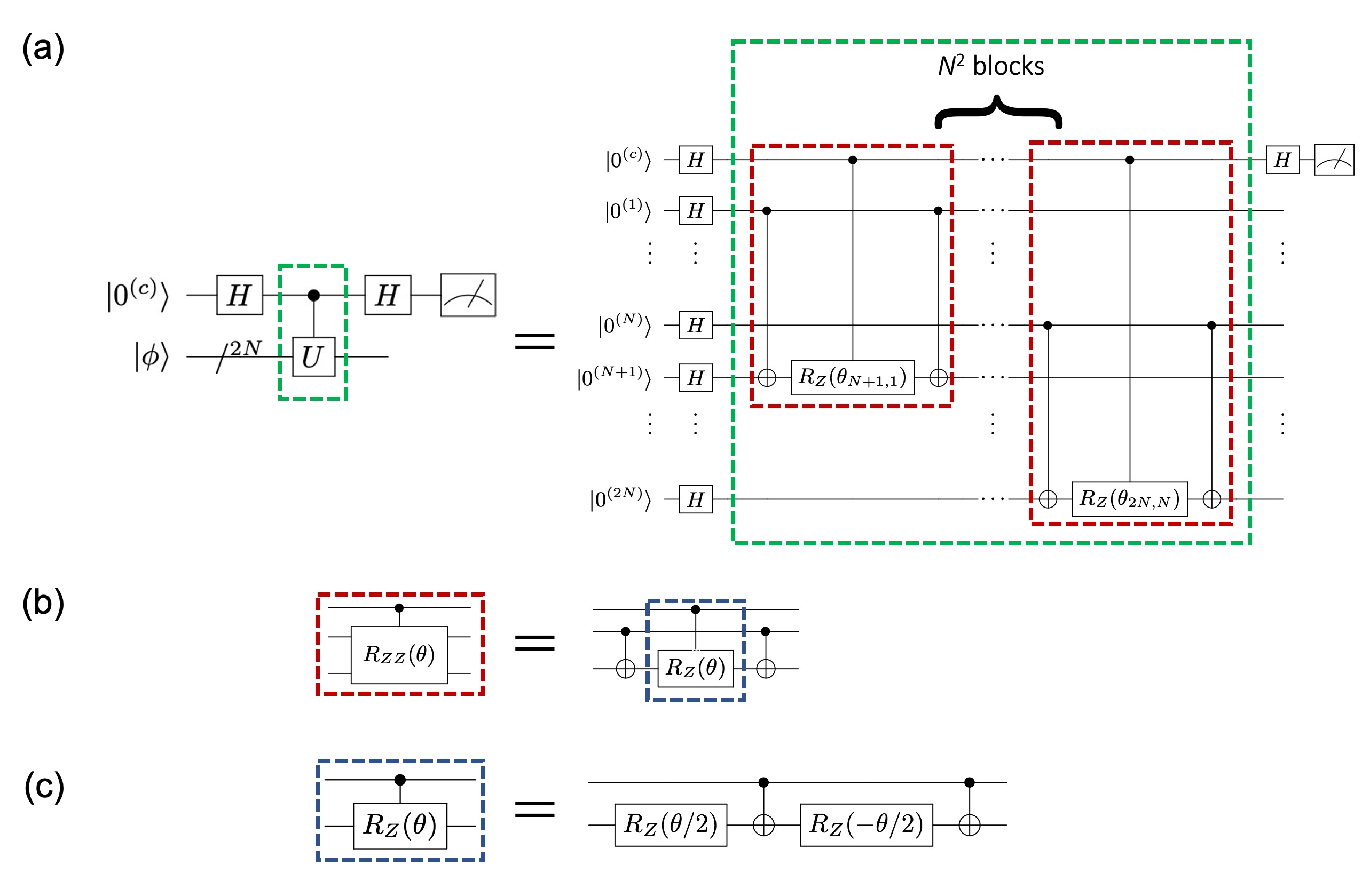}
\caption{ Quantum circuits for evaluating a real part of the overlap integral using the Hadamard test. (a) The Hadamard test circuit for the additive error estimation of the real part of the quantum overlap integral $\mathrm{Re}\braket{\phi\vert U\vert\phi}$. The real part of the quantum overlap integral can be obtained via the difference between the probabilities of getting $0$ and $1$ in the ancilla qubit (denoted as $c$) measurement. The detailed quantum circuit  implementation is presented: (b) the circuit synthesis of the controlled-$R_{ZZ}(\theta)=\exp(-\mathrm{i}\tfrac{\theta}{2} Z_{1}Z_{2})$ gate and   
    (c) the circuit synthesis of controlled-$R_{Z}(\theta)$, where $R_{Z}(\theta)=\exp(-\mathrm{i}\tfrac{\theta}{2} Z)$. 
    }
    \label{fig:quantumcircuits}
\end{figure}

Here, we present the quantum circuit synthesis for the additive error protocol based on the Hadamard test. 

The time-propagator 
can be factorized into two-qubit gates as 
\begin{align}
    U(A';t/2)=
    (-1)^{N}\prod_{p,q=1}^{N} \exp\left(-\mathrm{i}A'_{pq}\frac{t}{2}Z_{N+p}Z_{q}\right), 
\end{align}
where the two-qubit gate is a simple Ising $ZZ$-gate $R_{ZZ}^{(p,q)}(\theta_{pq})=\exp(-\mathrm{i}\tfrac{\theta_{pq}}{2} Z_{N+p}Z_{q})$. 
We estimate the overlap integrals using the Hadamard test method in Fig.~\ref{fig:quantumcircuits}(a), and the required controlled-$U$ gate, $V$, is decomposed as a simple product of controlled-$R_{ZZ}$ gates as   
\begin{align}
&V(A';\Delta t/2)= \nonumber \\
&\prod_{p,q=1}^{N}
(\ket{0^{(c)}}\bra{0^{(c)}}+\ket{1^{(c)}}\bra{1^{(c)}}\otimes R_{ZZ}^{(p,q)}(A'_{pq} t)), 
\end{align}
where $(c)$ denotes the control ancilla qubit. Accordingly, Fig.~\ref{fig:quantumcircuits} shows the detailed quantum circuit corresponding to the Hadamard test,  
where $A'=A+\pi t^{-1} I$ and $I$ is an identity matrix. We used the following identity for the parity operator,  $(\bigotimes_{k=1}^{2N}H_{k})P(\bigotimes_{k=1}^{2N}H_{k})=P_{Z}=(-1)^{N}\bigotimes_{k=1}^{N}\exp\left(-\frac{\pi}{2}\complexi Z_{k}Z_{N+k}\right)$, 
 to integrate it in the unitary propagator $U$ of the quantum Ising Hamiltonian. 
After all, it requires $2N+1$ qubits with $\mathcal{O}(N^{2})$ circuit depth and CNOT gates per single overlap integral evaluation and $\mathcal{O}(1/\epsilon^{'2})$ samples. 

\section{Alternative proof of Eq.~\eqref{eq:proofgaussianaverage}}
\begin{align}
&\mathcal{E}\left(\langle \mathbf{0}\vert
    \exp(\beta\mathcal{H}(W)) \vert \mathbf{0}\rangle\right)\nonumber \\
&=
    \mathcal{E}\left(\langle \mathbf{0}\vert    
\exp(\beta\sum_{j,k=1}^{N}W_{jk}X_{N+j}X_{k}) \vert \mathbf{0}\rangle\right)\nonumber \\
&=
    \mathcal{E}\left(\langle \phi\vert    
\exp(\beta\sum_{j,k=1}^{N}W_{jk}Z_{N+j}Z_{k}) \vert \phi\rangle\right)\nonumber\\
&=\frac{1}{2^{2N}}\sum_{\vec{x},\vec{x}'\in \{-1,1\}^{N} }
    \mathcal{E}\left(
\exp(\beta\sum_{j,k=1}^{N}W_{jk}x'_{j}x_{k}) \right)\nonumber\\
&=\frac{1}{2^{2N}}\sum_{\vec{x},\vec{x}'\in \{-1,1\}^{N} }
\prod_{j,k=1}^{N}\left[\frac{1}{\sqrt{2\pi}}\int_{-\infty}^{\infty}\mathrm{d}W_{jk}
\mathrm{e}^{-\frac{1}{2}W_{jk}^{2}+\beta W_{jk}x'_{j}x_{k}}\right]\nonumber\\
&=\exp\left(\frac{\beta^{2}N^{2}}{2}\right) , 
\end{align}
where we used $x_{j}^{'2}x_{k}^{2}=1$.

\section*{References}

%

\end{document}